# Growth and Strain Engineering of Trigonal Te for Topological Quantum Phases in Non-Symmorphic Chiral Crystals

**Rabindra Basnet** [1,2], **M. Hasan Doha** [1,2], **Takayuki Hironaka** [1,2], **Krishna Pandey** [3], **Shiva Davari** [1,2], **Katie M. Welch** [1,2], **Hugh O. H. Churchill** [1, 2*], **and Jin Hu** [1, 2,*]

[1] Department of Physics, University of Arkansas, Fayetteville, 72701, USA; rbasnet@uark.edu (R.B.); mdoha@uark.edu (M.H.D.); thironak@uark.edu (T.H.); sdavarid@uark.edu (S.D.); kmwelch@uark.edu (K.M.W.)
[2] Institute for Nanoscience and Engineering, University of Arkansas, Fayetteville, 72701, USA;
[3] Microelectronics-Photonics Graduate Program, University of Arkansas, Fayetteville, 72701, USA; kpandey@uark.edu
\* Correspondence: churchill@uark.edu (H.O.H.C.); jinhu@uark.edu (J.H.)

**Abstract:** Strained trigonal Te has been predicted to host Weyl nodes supported by a non-symmorphic chiral symmetry. Using low-pressure physical vapor deposition, we systematically explored the growth of trigonal Te nanowires with naturally occurring strain caused by curvature of the wires. Raman spectra and high mobility electronic transport attest to the highly crystalline nature of the wires. Comparison of Raman spectra for both straight and curved nanowires indicates a breathing mode that is significantly broader and shifted in frequency for the curved wires. Strain induced by curvature during growth therefore may provide a simple pathway to investigate topological phases in trigonal Te.

## 1. Introduction

Chiral materials possess structures with well-defined left- or right-handedness, which can be inverted by several symmetry operations, such as spatial inversion, mirror reflection, and roto-inversion. The structural chirality has been recognized to play an important role in diverse disciplines [1,2]. In condensed matter physics, the structural chirality of chiral crystals has been known to lead to a wide spectrum of electronic, magnetic, and optical phenomena, including chiral magnet and skyrmions [3,4], unusual optical and transport properties [5,6], etc. Recently, there has been growing interest in chiral crystals, owing to their topological nature. It has been predicted that the Kramers–Weyl nodes, which are locked at the time reversal invariant momenta (TRIM) in the Brillouin zone (BZ) and thus robust against annihilation [7], occur universally in all non-magnetic chiral crystals with spin–orbit coupling [4].

Among various chiral crystals, trigonal tellurium (Te) and Selenium (Se) are particularly interesting due to the potential to support highly tunable topological states. Trigonal Te crystallizes as helical chains with three atoms per turn along the *c*-axis (the screw axis), with chains ordered in hexagonal arrays within the *a*–*b* plane (Figure 1). Depending on the chirality of the screw axis, the space group can be $P3_121$ (right-handed) or $P3_221$ (left-handed). Although pristine Te at ambient pressure is a band semiconductor with an indirect gap of 0.3 eV [8], a transition to a strong topological insulator has been predicted for Te under shear strain [9]. Recent theoretical calculations have further revealed multiple pairs of Weyl nodes even in pristine Te and Se, which are protected by the non-symmorphic screw symmetry of a three-fold helical lattice [8,10]. Unlike the Kramers–Weyl nodes in chiral crystals mentioned above, the Weyl nodes in Te are not located at TRIM, so they display great tunability. For example, applying pressure on Te up to 2.17 GPa closes the band gap near the BZ *H* point (see Figure 1b for BZ for Te) and eventually creates multiple pairs of Weyl nodes near the Fermi level at the *K–H* line, leading to an inversion symmetry-breaking Weyl semimetal state [8]. These Weyl nodes are further movable along the *K–H* line with increasing pressure [8,10].

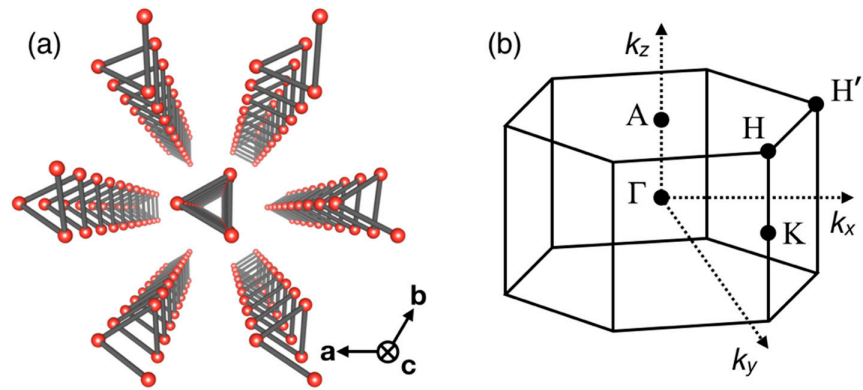

**Figure 1.** (**a**) Crystal structure and (**b**) first Brillouin zone (BZ) of trigonal tellurium (Te).

Unlike the tantalum arsenide (TaAs) family [11–21], which has been well-established as the inversion symmetry-breaking Weyl semimetals, the potential Weyl states in pristine and pressurized Te and Se, though predicted at almost the same time as the TaAs family, remain unexplored experimentally, with the exception of a recent report on angle-resolved photoemission spectroscopy for the valence band of bulk Te that reveals signatures consistent with Weyl nodes at and near the *H* point [22], and the possible quantum Hall effect of "massive" Weyl fermions [23]. The lack of reported progress investigating the topological properties of Te could originate in the experimental difficulty handling samples under hydrostatic pressure. In this work, we take advantage of the negative compression modulus along the *c*-axis in Te [24] to develop a pathway to investigate the potential Weyl states, as well as their tunability in curved Te nanowires. We show that, with the integration of strain and electrostatic gating, Te nanowires provide a versatile platform that will eventually allow electronic and spectroscopic characterization of the predicted Weyl states.

## 2. Materials and Methods

The Te nanowires used in this work were synthesized via a home-built physical vapor deposition (PVD) technique, as will be described in more detail later. Nanowires were characterized using polarized Raman spectroscopy (Horiba Jobin), scanning electron microscopy (SEM, FEI Nova Nanolab), and electronic transport. Raman spectra were collected using an excitation wavelength of 633 nm and an approximate power of 600 µW. Polarization of the excitation was adjusted with a half wave plate to be parallel to the *c*-axis of the nanowires (tangent to the curve for curved nanowires). Devices for electronic transport characterization were fabricated using a combination of electron beam lithography, thermal evaporation, and lift-off. A crucial fabrication detail to eliminate native oxide on the Te nanowires is a two-minute etch in 25% hydrochloric acid prior to contact deposition to remove a 4 nm thick native oxide. Transport measurements were voltage-biased DC current measurements.

## 3. Results

### 3.1. Nanowire Synthesis

As shown in Figure 1a, the chemical bonds in Te are highly anisotropic, with relatively strong bonds within the helical chains and relatively weak bonds between them. The degree of bonding anisotropy is intermediate between strongly layered and strongly 3D extremes, and these helical crystals have been described in the literature as both quasi-one-dimensional (1D) weakly bonded solids [25] and 1D van der Waals materials [26]. Anisotropic bonding in these materials is sufficient to permit synthesis of 1D nanowires [26–32] and two-dimensional (2D) flakes [33–38], as well as mechanical exfoliation of bulk crystals [39]. The relatively low melting (450 °C) and boiling points (988 °C) of Te permit 1D nanowires and 2D nano-flakes of Te to be easily synthesized via low-pressure physical vapor deposition (LP-PVD), as has been reported before [40,41]. In our experiments, the nanowire growth was conducted in a dual-heating zone tube furnace with the elemental source material (pure Te, 10 mg) and a growth substrate placed in hotter and colder zones, respectively, separated by 20 centimeters (Figure 2a,). At appropriate temperatures for both zones, the source vapor is carried by a flow of Ar gas from the hotter zone to colder zone, and condenses on the substrate to form nanostructures. In our growths, the Ar flow is turned on prior to the growth to flush the growth chamber, and turned off after growth ends, followed by a natural cool down of the furnace.

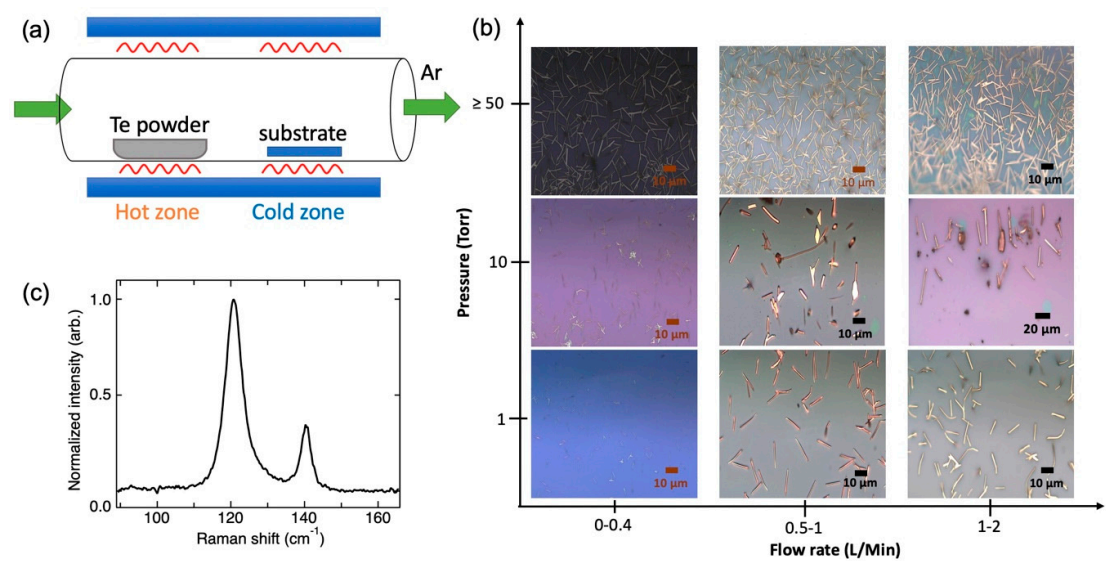

**Figure 2.** (**a**) Schematic of two-zone, low-pressure physical vapor deposition growth of Te nanowires. (**b**) Summary of pressure and flow rate dependence of growth outcomes. (**c**) Representative Raman spectrum of a synthesized Te nanowire.

With more than 100 growths with varying growth parameters, including temperature, Ar flow rate, chamber pressure, and growth duration, we mapped out the parameter space of growth conditions. We found that most of the successful growths occurred with the hot and the cold zones being kept at 450–550 °C and 200–350 °C, respectively. The density and morphologies of the nanostructures are sensitively dependent on the Ar flow rate and chamber pressure. As summarized in Figure 2b, high pressure clearly favors denser growth, while increasing the Ar flow rate leads to enhanced diameter of nanowires and even prompts the formation of 2D flakes.

In addition to the straight nanowires, curved wires appear when the chamber pressure is above 40 Torr. As will be discussed later, curved nanowires with built-in strain provide a versatile platform for electronic and spectroscopic characterization of the predicted Weyl states. Among various parameters, pressure provides a threshold of 40 Torr, above which curved nanowires start to form. Below this threshold, the growths regardless of any combination of other parameters are unable to fabricate curved nanowires. However, the curvature of nanowires appears not to be tunable with pressure. Above the threshold pressure (40 Torr), curved wires with varied curvatures can be obtained on the same substrate in each growth.

*3.2. Characterizations of Te Nanowires*

The electronic properties of the synthesized nanowires were examined by the standard two-terminal field-effect transistor devices (Cr/Au contacts) with a back gate. In Figure 3, we show the ambipolar, predominantly *p*-type transport of a typical device fabricated with a 160 nm-thick Te nanowire (Figure 3a), from which an on–off ratio of $8 \times 10^3$ was observed at 10 K. A contact resistance-limited, two-terminal field-effect mobility of 1000 cm$^2$/Vs can be extracted from the device transconductance shown in Figure 3b. The mobility is reduced with increasing temperature, reaching 200 cm$^2$/Vs at room temperature, which is comparable with other reports [27,42,43] and demonstrates the quality of the crystallinity of our Te nanowires. We estimated the mobility using the following equation [44]:

$$\mu = \frac{g_m L_G^2}{C V_{sd}} \quad (1)$$

where $g_m = \frac{dI_{sd}}{dV_g}$, $L_G$ is the gate length, C is the capacitance of the wire, and $V_{sd}$ is the bias voltage. Capacitance C was calculated by using Equation (2) [44],

$$C = \frac{2\pi\varepsilon L_G}{\ln\left[\frac{t_{ox} + a + \sqrt{(t_{ox}+a)^2 - a^2}}{a}\right]} \quad (2)$$

where $\varepsilon$ is the insulator dielectric constant, $t_{ox}$ is the insulator thickness, and a is the Te wire radius. Parameters for the device measured are $g_m = 4 \times 10^{-7}$ S, $L_G = 5$ μm, $V_{sd} = 50$ mV, $C = 4.8 \times 10^{-16}$ F, $\varepsilon = 3.9$, $t_{ox} = 300$ nm, and $a = 80$ nm.

Temperature-dependent resistance of a representative nanowire shows metallic behavior at zero backgate voltage, consistent with the strongly *p*-type character of the wires (Figure 3b), assuming Schottky-dominated contact resistance, which increases at low temperatures. The numerically calculated transconductance of the device at 10 K reaches a peak at a gate voltage of 45 V, followed by a minimum at lower gate voltages (Figure 3d). We speculate that this minimum originates from interband scattering, as trigonal Te has several bands adjacent to the valence band maximum [45]. We also observed transconductance oscillations near pinch-off, as shown in Figure 3d (above 45 V on the gate). Determining whether these oscillations originate from disorder, Coulomb blockade, or other mesoscopic mechanisms would require further investigation at lower temperatures.

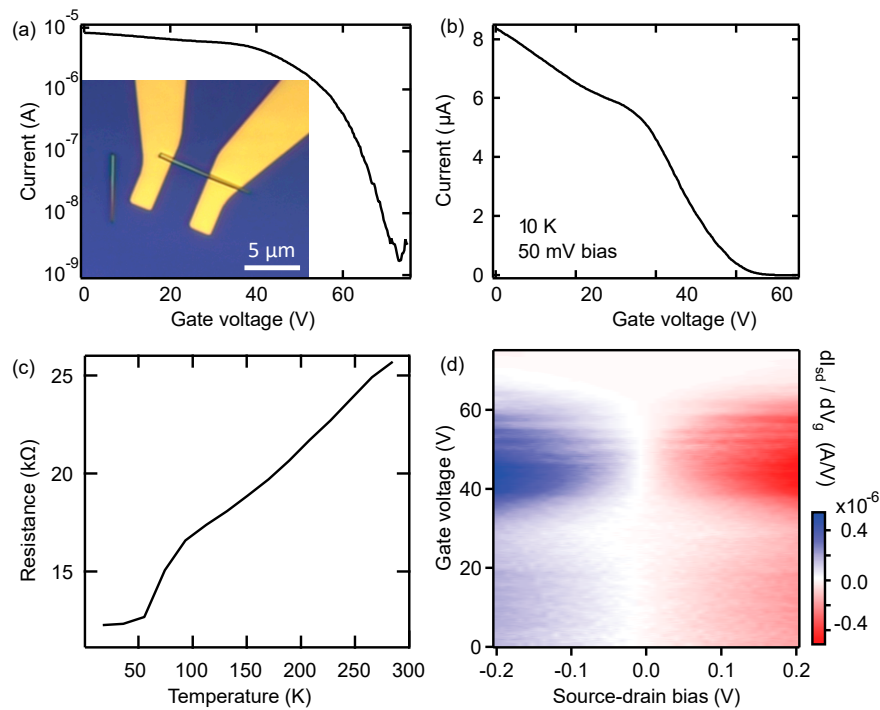

**Figure 3.** (**a**) Two-terminal Te nanowire device current at 50 mV source–drain bias and 10 K. *Inset:* Micrograph of a device. (**b**) Same data as (a) on a linear scale. (**c**) Temperature dependence of the nanowire resistance at zero gate voltage. (**d**) Numerically differentiated transconductance at 10 K.

*3.3. Strain Engineering of Te Nanowires*

As has been introduced above, Te nanowires are particularly interesting candidates that may host tunable inversion symmetry-breaking Weyl states [8,10]. However, the requirement of hydrostatic pressure [8,10] imposes a great challenge for the experimental characterization of Weyl states in these materials. Here, we introduce strain-engineering to manipulate the predicted Weyl phase in Te. The quasi-1D helical chain structure of Te generates a negative compression modulus along the *c*-axis (the chain direction) [24]; hence, applying hydrostatic pressure causes the helical Te chain to elongate along the *c*-axis, i.e., the uniaxial tensile strain along *c* analogizes the effect of hydrostatic pressure. In fact, before the concept of inversion symmetry-breaking Weyl state was fully established, tensile strain along *c* was predicted to close the band gap at the BZ H point for Te [9]. Such strain-induced gap closing [9] is indeed analogous to the scenario for the pressurized sample [8].

With this in mind, Te nanowires are particularly suitable for strain engineering, given the convenient strain application and large achievable strain in nanostructures as compared to bulk. Here, we demonstrate a new approach to apply strain in Te nanowires, the intrinsic built-in strain in curved nanowires.

Strain application using a flexible substrate has been widely reported for various nanostructures [46–49], including 2D Te flakes [35]. According to the first principles calculations, a transition to Weyl semimetal occurs at a pressure of 2.17 GPa, which corresponds to an elongation along the *c*-axis of less than 1% [24]. While the strain can be estimated from the bending of the substrate, Raman spectroscopy provides a convenient characterization of the strain effect. As shown in Figure 2c, the unstrained Te nanowire displays characteristic A1 and E2 Raman modes at 121 and 140 cm$^{-1}$ at room temperature, respectively, which agrees well with the previous reports for polarization parallel to the *c*-axis [50,51]. While mechanical bending works for both nanowires and the 2D flakes, one advantage of using nanowires is to take advantage of the built-in strain in curved wires, which provides an alternative and natural way to explore the possible Weyl state in Te and Se. As discussed before, during our synthesis efforts, curved Te nanowires with various curvatures can be obtained when chamber pressure is above a threshold pressure of 40 Torr. Tensile, compressive, and shear strains may all be present in a curved nanowire at different radial locations. Curvature radii down to 2 µm have been observed in our growths (Figure 4), typically among nanowires with smaller diameters. For a bent cylinder, the axial strain is $\varepsilon = x/\varrho$, where x is the radial distance from the center of the nanowire, and $\varrho$ is the radius of curvature of the wire (Figure 4c). We therefore expect, based on nanowire dimensions measured by SEM (Figure 4b), that strains of +/– a few percent should be achievable using naturally strained Te nanowires of this type.

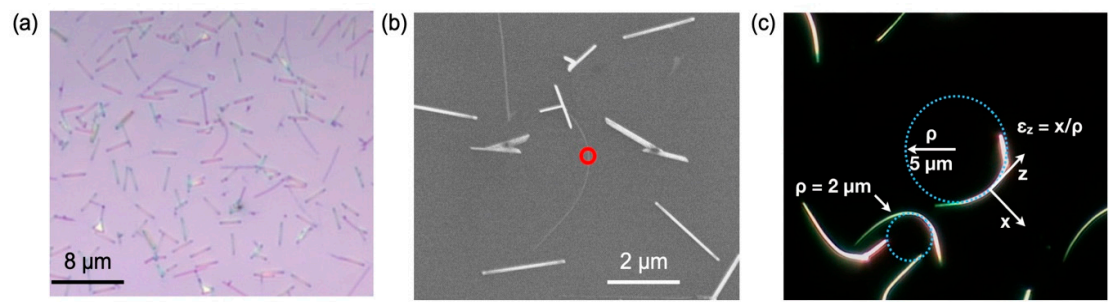

**Figure 4.** (**a**) Optical micrograph of physical vapor deposition (PVD)-grown Te nanowires, including small diameter curved nanowires. (**b**) Scanning electron microscopy (SEM) image of the same area shown in (a). The red circle indicates the approximate spot size and location of Raman spectroscopy

laser. (**c**) Darkfield optical micrograph of a different Te nanowire growth. Examples of different curvature radii are shown.

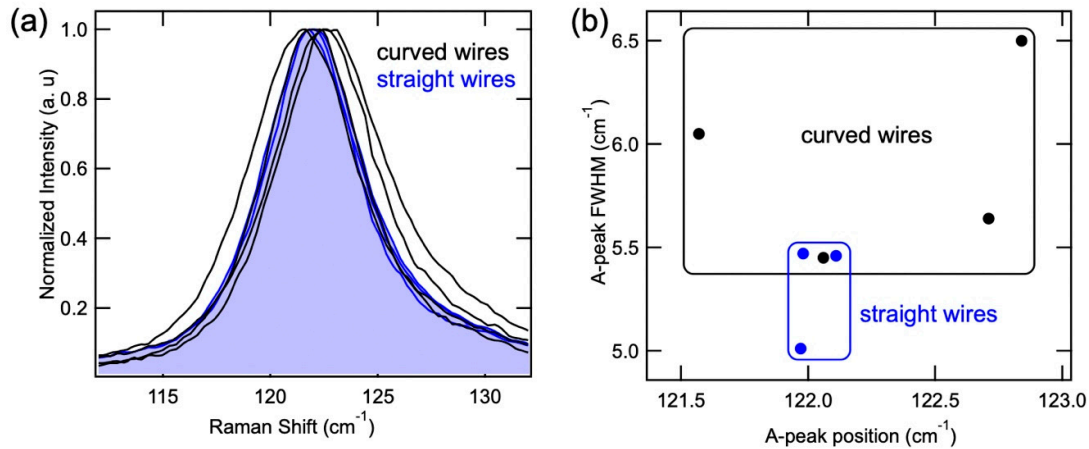

**Figure 5.** (**a**) Raman spectra of four curved (black) and three straight (blue) Te nanowires. (**b**) Widths and peak positions extracted from Lorentzian fits to the seven spectra shown in (a).

Optical and SEM images of a representative curved Te nanowire are shown in Figure 4a, b. We obtained indirect characterization of the strain state of this and three similar curved Te nanowires from the same growth using Raman spectroscopy (Figure 5a). As expected, both uniaxial strain and hydrostatic pressure have been shown previously to significantly shift the position of Raman active modes in Te [35,52]. Focusing on the most intense mode in the Raman spectrum of Te, the $A_1$ breathing mode, we find that straight Te nanowires display strongly uniform peak positions and widths (Figure 5b), while curved nanowires on average have broader peaks and significant shifts up and down in frequency, relative to straight wires. These results are consistent across measurements of four curved and three straight nanowires from the same growth on the same substrate.

To estimate the significance of frequency shifts of approximately 0.5 cm$^{-1}$, we note that hydrostatic pressure of 4 GPa shifts the $A_1$ mode of Te lower by about 10 cm$^{-1}$ [52], while the same mode shifts lower with *c*-axis uniaxial strain of 1% by 1 cm$^{-1}$ [35]. Thus, the shifts observed in our experiment might correspond to equivalent pressures of approximately 0.2 GPa or strains of 0.5%. Although a pressure an order of magnitude larger is expected to be necessary to achieve a transition to the Weyl semimetal state in Te, we note that our diffraction-limited spectroscopy samples the entire cross-section of the nanowire, which is expected to have regions of both compressive and tensile strain. Therefore, the most strained portions of the wire on the inner and outer edges may not be well represented in our spatially averaged spectra. We speculate that the fact that both redshifts and blueshifts were observed may be caused by the comparable length scales of the diffraction-limited laser spot and the nanowire diameters: Random misalignment of the beam could preferentially sample the tensile strained outer edge or the compressive strained inner edge of the nanowire. Future experiments using high-resolution tip-enhanced Raman spectroscopy will be required to clarify this point.

To clarify the predicted evolution of topological electronic states, a systematic study on Te nanowires with various diameter and curvature is needed. As mentioned above, among various parameters in our PVD growths, the chamber pressure appears to act as a switch turning on the curved wire formation at 40 Torr, but pressure is less efficient in fine-tuning the diameter and curvature of nanowires. On the other hand, since the diameter of the nanowires is highly dependent on the flow rate of the carrier gas (Figure 2b), fine-tuning of the flow rate could provide a control knob for the strain in nanowires, which will be investigated in future studies.

**4. Discussion: Experimental Examination of the Weyl State in Strained Te Nanowires**

The predicted strain-induced Weyl semimetal phase can be conveniently examined by magneto-transport measurements. Established common transport evidence [7,53,54] of a Weyl state includes the non-trivial Berry phase, which can be obtained from quantum oscillation [53], the planar Hall effect (PHE) [55–57], and the negative magnetoresistance (MR) induced by the chiral anomaly [56,58–60]. In topological materials, the non-trivial band topology gives rise to a non-trivial Berry phase, which shifts the phase of quantum oscillations. Quantum oscillations on unstrained bulk [61,62] and 2D Te [35] have been reported under a moderate magnetic field. From the Landau fan diagram established from the Shubnikov–de Haas quantum oscillation in MR, however, trivial Berry phase has been extracted [63], which does not support the existence of topological band. This could be understood in terms of the fact that the Weyl points predicted in the pristine Te are deep below the Fermi level and may not contribute to transport [8]. However, like pressurized Te [8], the strained sample should display Weyl points at the Fermi level, which would give rise to non-trivial Berry phase that can be probed in quantum oscillation.

One practical problem of the quantum oscillation experiment in nanowires is the size effect that may prohibit the formation of complete cyclotron orbits. From semiclassical theory, the cyclotron radius of the n$^{th}$ Landau level is given by $r_n = l_B \sqrt{n}$, where $l_B = \sqrt{\hbar/eB}$ is the magnetic length. For unpressurized pristine Te, n reaches 8 at B = 10 T [63]. Assuming the nanowire is clean enough that Landau levels are sufficiently sharp to be distinguished from each other, a minimum diameter of 23 nm is required to observe quantum oscillations at 10 T. It is worth noting that this value is expected to change in strained samples, because the size of Fermi surface varies with strain, which alters the quantum oscillation frequency and consequently changes the Landau level index n for given B.

The PHE describes the finite transverse Hall signal, even with the in-plane magnetic field, which is another consequence of non-trivial Berry phase and has been probed in various Dirac and Weyl semimetals [53,55–57,64,65]. Because a Hall bar geometry is needed to contact the sample, it is more suitable for 2D flakes rather than nanowires, with strain being applied by using the flexible substrate.

Compared to quantum oscillations and PHE, the chiral anomaly is more feasible and particularly applicable for nanowire samples. Chiral anomaly describes non-conservation of chiral charges, which originates from charge pumping between two Weyl cones with opposite chirality under parallel electric and magnetic fields. Such a phenomenon leads to negative longitudinal magnetoresistance, which has been claimed to exist in various topological materials [7,53,54]. However, a classical effect, current jetting, can also lead to negative longitudinal MR. In electron transport, once large conductance anisotropy is present, equipotential lines are strongly distorted, and thus, the current forms "jets", which could lead to negative longitudinal MR, particularly for asymmetric point-like electrical contacts and irregular sample shape [66,67]. For nanowires, however, current jetting should be minimized because of the large aspect ratio and symmetric voltage contacts fabricated by lithography. Therefore, the observation of negative longitudinal MR in strained, non-magnetic Te nanowires would provide strong support for the presence of topological non-trivial bands near the Fermi level.

In addition to strain, electrostatic gating provides another tuning parameter that is particularly useful to fine-tune the Fermi level to approach Weyl points. Gating has already been demonstrated in a previous section on unstrained, semiconducting nanowires. For the much more metallic Weyl semimetal phase in strained nanowires [6], effective gating is still possible with thin wires or with ionic liquid gating, as has been demonstrated in other metallic systems [68–70].

**Author Contributions:** Growth, R.B.; K.P.; and T.H.; Raman, T.H.; H. D. and H.C; Device Fabrication and characterization, T.H.; S.D.; K.W, and H.C.; writing, H.D.; R.B.; H.C.; and J.H.; supervision, H.C. and J.H..

**Funding:** J. H. acknowledges support from the U.S. Department of Energy (DOE), Office of Science, Basic Energy Sciences program under award DE-SC0019467 (personnel, growth, and material characterizations), H.C. is supported by DOE award number DE-SC0019467 (material characterization) and NSF award number DMR-1841821 (device fabrication). K.W. acknowledges support from NSF award number EEC-1757979.

**Conflicts of Interest:** The authors declare no conflict of interest.